%% file: ms.tex
\documentclass[12pt,preprint]{aastex}

\shorttitle{M13 Abundances}
\shortauthors{Johnson \& Pilachowski}

\begin{document}

\title{Oxygen and Sodium Abundances in M13 (NGC 6205) Giants: Linking Globular
Cluster Formation Scenarios, Deep Mixing, and Post--RGB Evolution}

\author{
Christian I. Johnson\altaffilmark{1,3,4} and
Catherine A. Pilachowski\altaffilmark{2} 
}

\altaffiltext{1}{Department of Physics and Astronomy, UCLA, 430 Portola Plaza,
Box 951547, Los Angeles, CA 90095-1547, USA; cijohnson@astro.ucla.edu}

\altaffiltext{2}{Department of Astronomy, Indiana University, Swain West 319, 
727 East Third Street, Bloomington, IN 47405-7105, USA; 
catyp@astro.indiana.edu}

\altaffiltext{3}{Visiting Astronomer, Kitt Peak National Observatory, National 
Optical Astronomy Observatories, which is operated by the Association of 
Universities for Research in Astronomy, Inc. (AURA) under cooperative agreement
with the National Science Foundation.  The WIYN Observatory is a joint facility
of the University of Wisconsin--Madison, Indiana University, Yale University, 
and the National Optical Astronomy Observatory.}

\altaffiltext{4}{National Science Foundation Astronomy and Astrophysics
Postdoctoral Fellow}

\begin{abstract}

We present O, Na, and Fe abundances, as well as radial velocities, for 113 red 
giant branch (RGB) and asymptotic giant branch (AGB) stars in the globular
cluster M13.  The abundances and velocities are based on spectra obtained with
the WIYN--Hydra spectrograph, and the observations range in luminosity from the
horizontal branch (HB) to RGB--tip.  The results are examined in the context of
recent globular cluster formation scenarios.  We find that M13 exhibits many 
key characteristics that suggest its formation and chemical enrichment are 
well--described by current models.  Some of these observations include: the 
central concentration of O--poor stars, the notable decrease in [O/Fe] (but 
small increase in [Na/Fe]) with increasing luminosity that affects primarily 
the ``extreme" population, the small fraction of stars with halo--like 
composition, and the paucity of O--poor AGB stars.  In agreement with recent 
work, we conclude that the most O--poor M13 giants are likely He--enriched and 
that most (all?) O--poor RGB stars evolve to become extreme HB and 
AGB--manqu{\'e} stars.  In contrast, the ``primordial" and ``intermediate" 
population stars appear to experience standard HB and AGB evolution.

\end{abstract}


\section{INTRODUCTION}

For many years globular clusters (GCs) were viewed as prototypical simple 
stellar populations containing stars of a single age and chemical composition.
However, a detailed examination of GC chemistry revealed
large star--to--star abundance variations of the light elements carbon
through aluminum (e.g., see reviews by Kraft 1994; Gratton et al. 2004; 2012).
While the anticorrelated behavior of carbon and nitrogen with increasing 
luminosity along the red giant branch (RGB), attributed to first dredge--up
(e.g., Iben 1965) and ``canonical extra mixing" (e.g., Denissenkov \& 
VandenBerg 2003), was observed in both cluster and field stars, a peculiar
pattern of enhanced N, Na, and Al abundances coupled with depleted O and Mg 
seemed to be only found in some cluster stars.  The simultaneous 
anticorrelation of O and Mg with Na and Al pointed to high temperature 
proton--capture burning as the likely source.  Unfortunately, it was not 
immediately clear if the processed material found in the photospheres of 
GC RGB stars was due to \emph{in situ} mixing or pollution from
a previous generation of more massive stars.

The comprehensive GC abundance survey by Carretta et al. 
(2009a,b) verified that these light element ``anomalies", in particular the 
O--Na anticorrelation, are likely present in \emph{all} Galactic GCs.
Additionally, several authors have now shown that the large 
star--to--star light element abundance variations found on the RGB are also 
present along the lower RGB, subgiant branch (SGB), and main--sequence turn off
(e.g., Gratton et al. 2001; Cohen \& Mel{\'e}ndez 2005; Bragaglia et al. 2010).
This observation indicates that the unique abundance patterns of GCs are the 
result of the cluster formation and subsequent evolution rather than \emph{in 
situ} processing.  Recent photometric observations have discovered that many 
(all?) GCs exhibit multiple evolutionary sequences in their color--magnitude 
diagrams (e.g., see reviews by Piotto 2009; Gratton et al. 2012).  Since most 
clusters exhibit a $<$0.1 dex spread in [Fe/H]\footnote{[A/B]$\equiv$log(N$_{\rm A}$/N$_{\rm B}$)$_{\rm star}$--log(N$_{\rm A}$/N$_{\rm B}$)$_{\sun}$ for elements A and B.} (e.g., Carretta et al. 2009c), except for a few notable cases with
significant [Fe/H] dispersion, the multiple photometric sequences are believed 
to be driven by He abundance differences.  While the source of He, and 
subsequently the light element variations, is not known, plausible candidates 
include: $\sim$5--9 M$_{\rm \odot}$ asymptotic giant branch (AGB) stars (e.g., 
Ventura \& D'Antona 2009; 2011), rapidly rotating massive main--sequence stars 
(e.g., Decressin et al. 2010), and massive binary stars (e.g., de Mink et al. 
2009). 

Although recent GC formation models incorporating winds from 
intermediate mass and massive stars mixed with pristine gas are able to 
reproduce many of the light element abundance trends unique to the cluster 
environment (e.g., Decressin et al. 2010; Valcarce \& Catelan 2011; D'Ercole 
et al. 2012), the very low oxygen abundances ([O/Fe]$<$--0.4) found in some 
GC RGB stars seems to require additional processing.  While the old 
paradigm that the O--Na anticorrelation is entirely driven by \emph{in situ} 
deep mixing in cluster RGB stars is clearly incorrect, the discovery that many 
GC stars are also He--rich has an important consequence for 
resurrecting a modified deep mixing scenario.  D'Antona \& Ventura 
(2007) showed that it is possible for a metal--poor star that is both He--rich 
\emph{and} initially moderately O--poor ([O/Fe]$\sim$--0.2) and Na--rich to 
further deplete oxygen down to [O/Fe]$\sim$--1, without a significant
change in the [Na/Fe] ratio.

In light of this, M13 is a particularly illuminating case.  It has long been
known that M13 hosts some of the most O--poor and Na/Al--rich RGB stars of any 
cluster, and that these stars appear to be found preferentially near the 
RGB--tip (e.g., Kraft et al. 1992; Pilachowski et al. 1996; Kraft et al. 1997; 
Cavallo \& Nagar 2000; Sneden et al. 2004; Cohen \& Mel{\'e}ndez 2005; Johnson 
et al. 2005).  However, the sample size of stars for which [O/Fe] has been 
determined is $\sim$5 times less than that for which [Na/Fe] and [Al/Fe] have 
been measured.  Since the [O/Fe] ratio may be the most sensitive 
indicator of deep mixing (D'Antona \& Ventura 2007), in this work we have 
measured [O/Fe] (and [Na/Fe]) abundances for $>$100 RGB and AGB stars ranging 
in luminosity from the RGB bump to the RGB--tip.  We now use this extended 
sample to examine how M13's extreme O--Na anticorrelation extension fits into 
the modern picture of GC formation and evolution.

\section{OBSERVATIONS, DATA REDUCTION, AND ANALYSIS}

All observations for this project were obtained on 2011 May 19--20 using the 
WIYN 3.5m telescope instrumented with the Hydra multifiber positioner and bench
spectrograph.  A single spectrograph setup, with wavelength coverage ranging 
from 6050--6350 \AA, was used to obtain high signal--to--noise ratio 
(S/N$>$100), moderate resolution (R$\approx$18,000) spectra of 113 RGB and AGB 
stars.  A color--magnitude diagram illustrating the evolutionary state of the 
observed stars is shown in Figure \ref{f1}.  All coordinates, optical 
photometry, and membership probabilities were taken from Cudworth \& Monet 
(1979).  Infrared photometry was taken from the 2MASS database (Skrutskie et 
al. 2006).  To ensure membership, we only observed targets with P$>$70$\%$.

The data reduction and analysis closely follow the techniques outlined in 
Johnson et al. (2005) and Johnson \& Pilachowski (2010).  To briefly 
summarize, effective temperature (T$_{\rm eff}$) and surface gravity (log(g))
were set for each star using dereddened V and K$_{\rm S}$ photometry.  We 
initially assumed [Fe/H]=--1.50 and a microturbulence (vt) value of 2 km 
s$^{\rm -1}$ and interpolated within the $\alpha$--enhanced, AODFNEW grid of 
ATLAS9 model atmospheres (Castelli et al. 1997).  The final model metallicity
was set as the average [Fe/H] derived from Fe I and Fe II lines, and vt was 
set by removing any trend in Fe I abundance versus line strength.

Abundances for Fe I and Fe II were derived via equivalent width measurements
while O and Na abundances were determined through spectrum synthesis.  All
abundances were calculated using the 2010 version of the LTE line analysis code
MOOG (Sneden et al. 1973).  The linelist was the same as that used in Johnson
\& Pilachowski (2010).  A summary of all derived model atmosphere 
parameters, coordinates, abundances, and radial velocities is provided in 
Table 1.

\section{BASIC RESULTS}

Despite exhibiting large light element abundance variations, M13 has always 
been characterized by a single metallicity ([Fe/H]$\sim$--1.5).  We find in
agreement with past large sample studies (e.g., Pilachowski et al. 1996; 
Sneden et al. 2004; Cohen \& Mel{\'e}ndez 2005) that M13 is moderately 
metal--poor and exhibits small star--to--star variation in [Fe/H].  In 
particular, we find $\langle$[Fe/H]$\rangle$=--1.57 ($\sigma$=0.07), with an 
average agreement between [Fe/H] derived by Fe I and Fe II (in the sense 
[FeI/H]--[FeII/H]) of --0.04 ($\sigma$=0.15).  The Fe I abundances are based
on an average of 27 lines ($\sigma$=4), with a typical line--to--line
dispersion of 0.13 dex ($\sigma$=0.02).  In contrast, the Fe II abundances are
based on 1--3 lines, with an average line--to--line dispersion of 0.11 dex
($\sigma$=0.07).  While the agreement between [FeI/H] and [FeII/H] is good 
for most stars, Table 1 shows that there is some disparity for the coolest,
most luminous giants.  This is likely due to a combination of mass loss, 
variability (if V$\la$12.5; Kopacki et al. 2003), model atmosphere 
deficiencies, and NLTE effects.

\begin{figure}
\epsscale{1.00}
\plotone{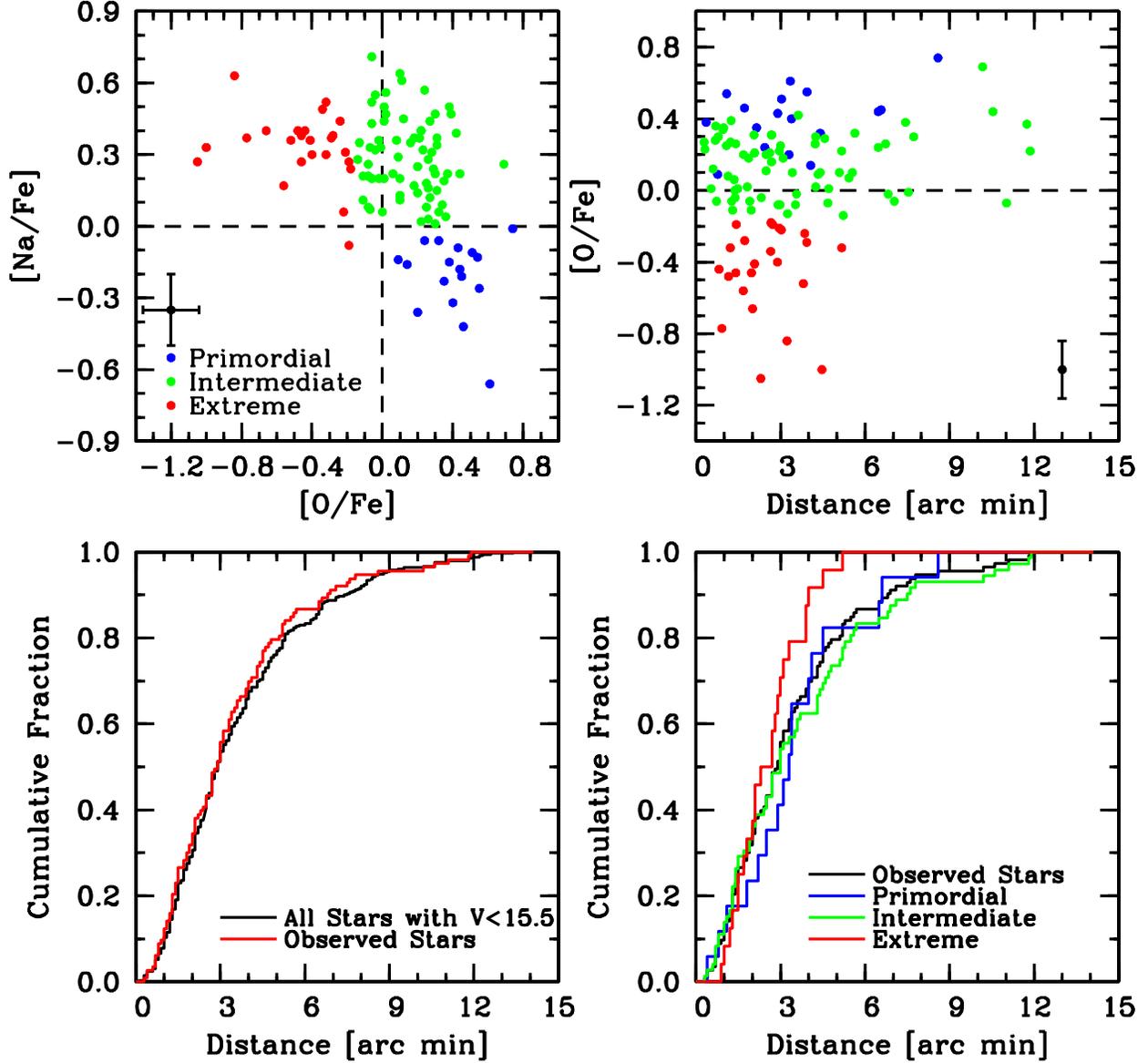}
\caption{The top left panel plots [Na/Fe] versus [O/Fe], and the top right 
panel shows [O/Fe] versus distance from the cluster center.  The bottom left 
panel illustrates the cumulative fraction as a function of radial distance for
all proper motion members (solid black line) and our observed distribution 
(solid red line).  The bottom right panel compares the cumulative distribution 
of our observations based on composition.}
\label{f2}
\end{figure}

In agreement with past work (e.g., Pilachowski et al. 1996; Kraft et al. 1997;
Cavallo \& Nagar 2000; Sneden et al. 2004; Cohen \& Mel{\'e}ndez 2005;
Johnson et al. 2005), we find large star--to--star abundance variations for 
both [O/Fe] and [Na/Fe] and reproduce the well--known O--Na anticorrelation 
(see Figure \ref{f2}).  The [O/Fe]\footnote{Note that we measured [O/Fe] 
relative to [Fe/H]$_{\rm avg.}$ rather than [FeII/H].} ratio ranges from --1.05
to $+$0.74, with an average [O/Fe]=$+$0.06 ($\sigma$=0.34).  Similarly, [Na/Fe]
ranges from --0.66 to $+$0.71, with an average of [Na/Fe]=$+$0.23 
($\sigma$=0.24).  For Na, the average measurement error is 0.08 dex 
($\sigma$=0.07).  While the O abundance is derived solely from the 6300 \AA\ 
[O I] line, the typical synthesis fitting uncertainty is $\la$0.1 dex.

\begin{figure}
\epsscale{1.00}
\plotone{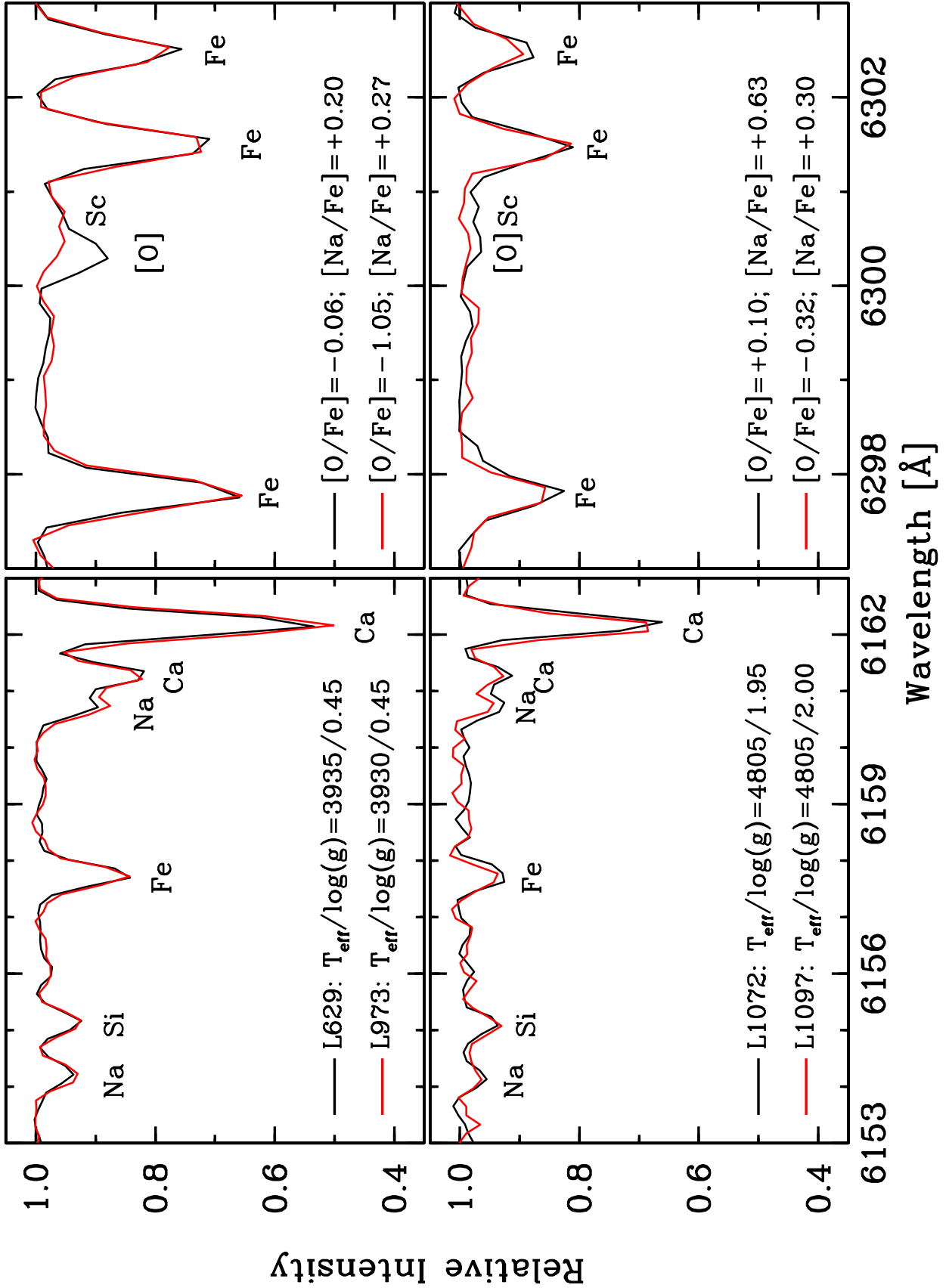}
\caption{Sample spectra for two sets of stars with similar T$_{\rm eff}$/log(g)
but different [O/Fe] and [Na/Fe] abundances.}
\label{f3}
\end{figure}

We did not apply any NLTE corrections to the [Na/Fe] abundances.  Although
departures from LTE are expected for cool, metal--poor giants, the magnitude 
of the corrections likely do not exceed $\sim$0.1 dex for the 6154/6160 \AA\
Na I lines (e.g., Lind et al. 2011).  The results presented here indicate
a correlation between luminosity and the O--Na anticorrelation, and it is 
important to ensure this result is not purely a consequence of NLTE effects or 
model atmosphere deficiencies.  While it is difficult to completely rule out
these effects, we note that: (1) abundance analyses of evolved RGB stars in
the similar metallicity GC M3 do \emph{not} find a correlation between O/Na 
abundance and luminosity (e.g., Sneden et al. 2004; Cohen \& Mel{\'e}ndez 2005)
and (2) as can be seen in Figure \ref{f3} there is a clear variation in O/Na
line strength among stars of similar luminosity.

\begin{figure}
\epsscale{1.00}
\plotone{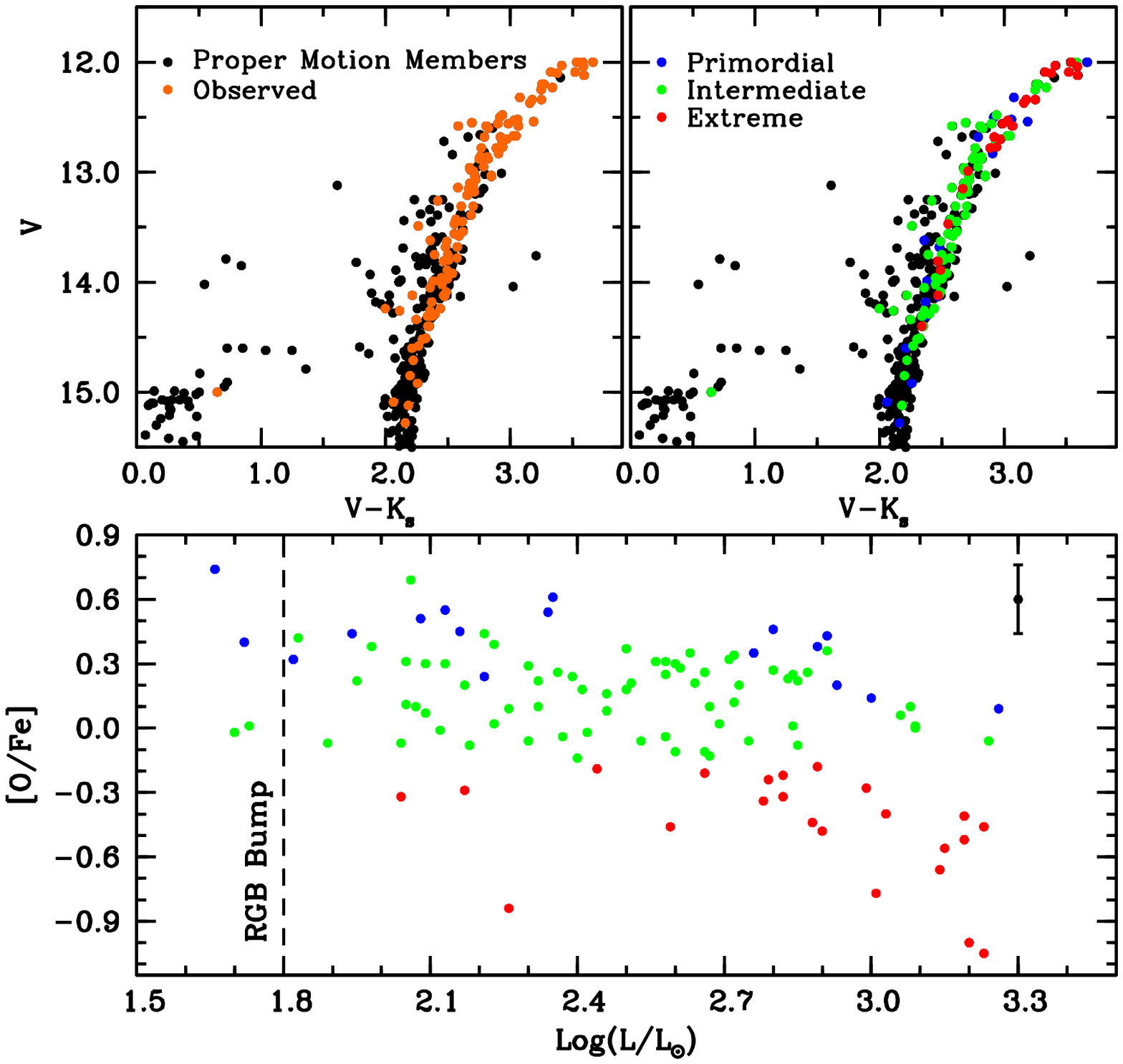}
\caption{The top left panel shows a V versus V--K$_{\rm S}$ color--magnitude
diagram indicating the stars observed for this program.  The top right panel 
shows the same stars differentiated by chemical composition (see also $\S$3). 
The bottom panel plots [O/Fe] versus log(L/L$_{\rm \odot}$).  The different 
color symbols have the same meaning as in the top right panel.}
\label{f1}
\end{figure}

Following the typical naming scheme for GC sub--populations 
(e.g., Carretta et al. 2009a), in Figures \ref{f1}--\ref{f2} we differentiate 
M13 stars into the ``primordial", ``intermediate", and ``extreme" populations 
based on their [O/Fe] and [Na/Fe] abundances.\footnote{Note that our 
definitions are slightly different than those used in previous studies.  Here 
we designate extreme stars by [O/Fe]$<$--0.15, primordial stars by 
[Na/Fe]$<$$+$0.00, and the remainder as intermediate stars.}  We find that the 
primordial, intermediate, and extreme populations constitute 15$\%$, 63$\%$,
and 22$\%$ of our sample, respectively, which is typical for Galactic 
GCs (e.g., Carretta et al. 2009a).  Interestingly, as can be 
seen in Figures \ref{f1}--\ref{f2}, the extreme population seems to 
differentiate itself by consisting of stars predominantly near the RGB--tip,
a result noted in many past studies (e.g., Kraft et al. 1997), and is the 
most centrally concentrated.  Additionally, we note that none of the AGB stars
in our sample are particularly O--poor (see also Pilachowski et al. 1996).  We 
discuss the implications of these observations further in $\S$4.

In addition to determining abundance ratios, we also measured radial velocities
for all stars using the IRAF \emph{fxcor} routine.  We find an average 
heliocentric radial velocity (RV) of --244.7 km s$^{\rm -1}$ ($\sigma$=6.1), 
which is in good agreement with past studies (e.g., Lupton et al. 1987).  The
average measurement error is $\sim$0.2 km s$^{\rm -1}$.  The small 
star--to--star velocity dispersion indicates that all of our observed targets 
are likely cluster members.  Interestingly, the extreme population exhibits an
average RV that is $\sim$2 km s$^{\rm -1}$ larger than the primordial and 
intermediate stars, which have identical average RVs.  However, this may be
due to the fact that most extreme population stars are near the RGB--tip
and therefore likely to be variables (e.g., Kopacki et al. 2003).

\section{DISCUSSION AND CONCLUSIONS}

As mentioned in $\S$1, recent models predict that GCs likely 
form in (at least) two distinct episodes.  In this scenario, the first star 
formation event produces stars with halo--like composition (the primordial 
population), and then the $\ga$5 M$_{\rm \odot}$ progeny of the first 
generation pollute the cluster with material heavily processed by high 
temperature proton--capture burning, including newly synthesized He.  This new 
material may be funneled to the cluster core (e.g., D'Ercole et al. 2008) where
the second generation stars (intermediate and extreme populations) form; 
however, it appears that some dilution with pristine gas is required to 
reproduce the observed light element abundance trends (e.g., Prantzos et al. 
2007; D'Ercole et al. 2011).  Some implications of this scenario are that: (1) 
the primordial population is preferentially stripped relative to the second 
generation stars, (2) the second generation stars may be significantly 
He--enhanced and more centrally concentrated, and (3) the extra He, in addition
to producing multiple evolutionary sequences in cluster color--magnitude 
diagrams, may cause some stars to experience \emph{in situ} deep mixing above 
the RGB bump and/or cause the most He--rich stars to become RGB--manqu{\'e}, 
AGB--manqu{\'e}, or extreme blue horizontal branch (HB) stars.  As we discuss 
below, M13 appears to exhibit many of these 
characteristics.\footnote{Interestingly, multiple sequences in M13 
color--magnitude diagrams have yet to be found (see Sandquist et al. 2010 for
a recent update.)}

\subsection{Supporting Observations of Globular Cluster Formation Models}

We noted in $\S$3 that the primordial population constitutes a 
considerably smaller fraction of stars in M13 (15$\%$) than the intermediate 
and extreme populations.  This is consistent with the cluster formation 
scenario mentioned above where a significant percentage (up to $\sim$90$\%$) of
first generation but not second generation stars are lost early in the 
formation process.  Although our estimate is somewhat lower than the 34$\%$ 
primordial fraction determined by Carretta et al. (2009a; their Table 5), we 
note that there is typically no clear separation between the primordial
and intermediate populations.  However, the dominance of the intermediate
population in M13 strongly suggests that its formation and chemical
enrichment followed the same path as other halo GCs.

Similarly, we show in Figure \ref{f2} that the extreme population appears to 
be marginally more centrally concentrated than the primordial and intermediate
stars.  This is supported by the results of two--sided KS tests, which 
indicate that the primordial and intermediate populations trace the same radial
distribution (KS--prob=0.9018) but the extreme population is different than
both (KS--prob$_{\rm P,E}$=0.1603; KS--prob$_{\rm I,E}$=0.0935).  Although
the statistical significance is marginal, we note that similar results have 
been found in a few other clusters where the central concentration of extreme 
stars is supported by independent observations of radial changes in the 
color--magnitude diagram (e.g., Carretta et al. 2010; Lardo et al. 2011).  
While the dynamical evolution of a GC is expected to smear out 
the radial profile and uniformly mix the various populations after a Hubble 
time (e.g., Decressin et al. 2010; but see also Bekki 2010), the fact that M13 
and other clusters still show a semblance of the extreme stars being centrally 
concentrated is evidence in support of current cluster formation models.  In 
this light, $\omega$ Cen is a particularly illustrative example.  Since the 
core relaxation time is similar to the cluster age, $\omega$ Cen likely 
preserves early formation history clues.  In fact, Johnson \& Pilachowski 
(2010) and Gratton et al. (2011) find a clear composition dependence on radial 
location, with the extreme stars being the most centrally concentrated and the 
primordial stars the least. 

\subsection{Connecting to the New Deep Mixing Model}

Although we now know that the historical argument relating \emph{in situ}
deep mixing and the O--Na anticorrelation is incorrect, a modified version has 
recently been resurrected to explain cluster RGB stars with [O/Fe]$\la$--0.4
(e.g., D'Antona \& Ventura 2007).  As mentioned in $\S$1, predicted yields from
both $>$5 M$_{\rm \odot}$ AGB and massive main--sequence stars generally fail 
to produce second generation stars with [O/Fe]$\la$--0.4 and thus a secondary 
process is required.  

Interestingly, our M13 observations (and those of past authors) appear to 
verify the predictions of the D'Antona \& Ventura (2007) model (see also Figure
\ref{f1}): (1) all of the known stars with [O/Fe]$\la$--0.4 are located well 
above the RGB bump, (2) in general there is a monotonic decrease in [O/Fe] 
with increasing luminosity for the extreme population, and (3) at the highest 
luminosity there is a large difference in $\langle$[O/Fe]$\rangle$ between the 
extreme and intermediate populations but only $\sim$0.1 dex increase in 
$\langle$[Na/Fe]$\rangle$ for the extreme stars.  We believe the requirements 
to induce deep mixing (enhanced He and initially low [O/Fe]) are also met for 
the extreme M13 giants.  

Figure \ref{f1} shows that (with one exception) the lowest [O/Fe] ratio found 
at log(L/L$_{\rm \odot}$)$<$2.8 is consistently at [O/Fe]$\sim$--0.3.  Note 
that this is consistent with the Cohen \& Mel{\'e}ndez (2005) observations 
that do not find stars below the RGB bump with [O/Fe]$<$--0.2.  This supports 
the idea that the low [O/Fe] values found only in the brightest M13 RGB stars 
is an evolutionary effect and that significant O--depletion does not occur at 
low RGB luminosities.  With regard to He--enhancement, we do not have direct 
He measurements for these stars but note that the most Na/Al--rich (and thus 
O--poor) stars in $\omega$ Cen (Dupree et al. 2011) and NGC 2808 (Pasquini et 
al. 2012) have enhanced He.  We also find ancillary evidence, similar to that
found by Carretta et al. (2006) in NGC 2808, in support of He--enrichment from
the increase in [Fe/H] from --1.58 in the intermediate population to --1.54 in
the extreme population.\footnote{We caution the reader on this point 
because the [Fe/H] difference is small and several of the extreme population
stars are known to be variable.}  However, we note that Sandquist et al. (2010)
does not find significant evidence for He--enrichment in M13.  On the other 
hand, if the O--poor stars are He--rich then the fact that deep mixing appears 
to be activated at a single luminosity (log(L/L$_{\rm \odot}$)$\sim$2.8) may be 
evidence in support of the extreme stars having a small He spread.  This is 
qualitatively in agreement with photometric studies that often find discrete 
populations in cluster color--magnitude diagrams rather than a spread (e.g., 
Piotto 2009).

\subsection{Composition and Post--RGB Evolution}

In the scenario described above, the most He--rich stars likely undergo
deep mixing that has the observational effect of decreasing [O/Fe]; however, 
it also increases the envelope He abundance to as much as Y=0.5 (e.g., D'Antona
\& Ventura 2007).  Since He--enhancement may be strongly manifest in HB and AGB
evolution, we can look to these stars for clues regarding He--enhancement and 
RGB evolution.  One of the most notable features of Figure \ref{f1}, which
has been shown previously with Na abundances (e.g., Pilachowski et al. 1996), 
is the lack of extreme stars on the AGB.\footnote{With the present 
data we are unable to differentiate AGB and RGB stars near the RGB--tip.  
However, we expect most, if not all, of the stars near the RGB--tip to be 
first ascent giants because of the short evolutionary timescale of AGB stars.}
Since we find the extreme stars to constitute $\sim$20$\%$ of M13's total 
population, we should expect to find $\sim$2--3 super O--poor AGB stars in 
our sample.  Interestingly, we find that only the primordial and intermediate 
AGB stars are present in about the same proportion as on the RGB.  Following 
similar results in other GCs (e.g., Norris et al. 1981; Campbell et al. 2010;
Gratton et al. 2010), we conclude that in M13 only the primordial and 
intermediate populations undergo standard HB and AGB evolution.

What about the fate of the extreme population?  M13 is known to contain a 
bimodal and extreme blue HB (e.g., see Sandquist et al. 2010 and references 
therein).  Circumstantial evidence supports the idea that the ``faint peak" 
population of HB stars, which have very high T$_{\rm eff}$, were also once the 
most O--poor giants on the RGB.  In particular, the fraction of faint peak 
relative to total HB stars is about equal to the fraction of extreme
to total RGB stars.  The faint peak stars were also found by Sandquist et al. 
(2010) to be more centrally concentrated than the ``intermediate" and ``bright 
peak" populations.  Furthermore, the Sandquist et al. (2010) data indicate 
that: (1) the fraction of AGB--manqu{\'e} to total AGB stars is $\sim$23$\%$ 
and (2) the origin of the AGB--manqu{\'e} stars is likely the bluest part of 
the HB.  Additionally, Peterson et al. (1995) provide [O/Fe] abundances for 
cool HB stars in M13 and do not find any with [O/Fe]$<$0.  All of these 
observations suggest that the extreme RGB stars evolve from the RGB to the 
bluest end of the HB and then become AGB--manqu{\'e} stars.

\subsection{Final Thoughts}

The results presented here have allowed us to re--examine M13 in light of 
recent advances in our understanding of GC formation.  M13 may be well 
explained by the new ``standard" picture in which first generation stars with 
halo--like composition are preferentially lost early in the cluster evolution, 
and a second, more enriched population forms in the cluster center from gas 
processed and ejected by $>$5 M$_{\rm \odot}$ first generation stars.  For M13,
this has the effect of instigating \emph{in situ} deep mixing in the most 
He--rich giants and perhaps causing them to terminate their evolution before 
ascending the AGB.  In fact, proper modeling of the warmest HB stars in 
clusters like M13 may require considering composition changes to the RGB 
envelope due to \emph{in situ} mixing.  However, two outstanding issues remain:
(1) will precise photometry in the inner part of the cluster finally reveal 
multiple populations and (2) are some M13 stars actually He--rich?

\acknowledgements

This material is based upon work supported by the National Science 
Foundation under award No. AST--1003201 to CIJ.  CAP gratefully acknowledges 
support from the Daniel Kirkwood Research Fund at Indiana University.

\clearpage

\clearpage

\clearpage

\clearpage
\input{tab1_full.tex}

\end{document}

%% file: tab1_full.tex
\tablenum{1}
\tablecolumns{16}
\tablewidth{0pt}

\begin{deluxetable}{cccccccccccccccc}
\tabletypesize{\scriptsize}
\rotate
\tablecaption{Basic Data and Results}
\tablehead{
\colhead{Star Name\tablenotemark{a}}       &
\colhead{Alt.}       &
\colhead{RA}	&
\colhead{DEC}	&
\colhead{V}       &
\colhead{K$_{\rm S}$}       &
\colhead{log(L/L$_{\rm \odot}$)}       &
\colhead{T$_{\rm eff}$}       &
\colhead{log(g)}       &
\colhead{[Fe/H]}       &
\colhead{vt}       &
\colhead{[FeI/H]}       &
\colhead{[FeII/H]}       &
\colhead{[O/Fe]}       &
\colhead{[Na/Fe]}       &
\colhead{RV}       \\
\colhead{}       &
\colhead{Name}       &
\colhead{J2000}	&
\colhead{J2000}	&
\colhead{}       &
\colhead{}       &
\colhead{}       &
\colhead{(K)}       &
\colhead{(cgs)}       &
\colhead{avg.}       &
\colhead{km s$^{\rm -1}$}       &
\colhead{}       &
\colhead{}       &
\colhead{}       &
\colhead{}       &
\colhead{km s$^{\rm -1}$}       
}

\startdata
L 324	&	 V11	&	250.402711	&	36.443214	&	12.00	&	8.465	&	3.23	&	3955	&	0.45	&	$-$1.50	&	2.50	&	$-$1.67	&	$-$1.32	&	$-$0.46	&	$+$0.27	&	$-$242.7	\\
L 598	&	\nodata	&	250.424834	&	36.447735	&	12.00	&	8.335	&	3.26	&	3895	&	0.40	&	$-$1.44	&	2.20	&	$-$1.57	&	$-$1.31	&	$+$0.09	&	$-$0.14	&	$-$257.6	\\
L 629	&	\nodata	&	250.427326	&	36.448975	&	12.00	&	8.418	&	3.24	&	3935	&	0.45	&	$-$1.57	&	2.15	&	$-$1.71	&	$-$1.42	&	$-$0.06	&	$+$0.20	&	$-$232.4	\\
L 194	&	 II$-$90	&	250.383343	&	36.474979	&	12.03	&	8.616	&	3.19	&	4015	&	0.50	&	$-$1.49	&	2.30	&	$-$1.53	&	$-$1.45	&	$-$0.41	&	$+$0.36	&	$-$239.3	\\
L 973	&	 I$-$48	&	250.462198	&	36.481819	&	12.04	&	8.452	&	3.23	&	3930	&	0.45	&	$-$1.50	&	2.35	&	$-$1.68	&	$-$1.32	&	$-$1.05	&	$+$0.27	&	$-$249.5	\\
L 835	&	 V15	&	250.445965	&	36.432686	&	12.09	&	8.763	&	3.14	&	4060	&	0.60	&	$-$1.50	&	2.15	&	$-$1.56	&	$-$1.43	&	$-$0.66	&	$+$0.40	&	$-$258.3	\\
L 954	&	 IV$-$25	&	250.459624	&	36.404362	&	12.09	&	8.568	&	3.19	&	3960	&	0.50	&	$-$1.48	&	2.30	&	$-$1.62	&	$-$1.34	&	$-$0.52	&	$+$0.36	&	$-$251.9	\\
L 940	&	\nodata	&	250.457090	&	36.463585	&	12.10	&	8.722	&	3.15	&	4030	&	0.55	&	$-$1.53	&	2.00	&	$-$1.62	&	$-$1.44	&	$-$0.56	&	$+$0.17	&	$-$246.5	\\
L  70	&	 II$-$67	&	250.348070	&	36.504818	&	12.12	&	8.527	&	3.20	&	3925	&	0.50	&	$-$1.45	&	2.35	&	$-$1.66	&	$-$1.24	&	$-$1.00	&	$+$0.33	&	$-$241.8	\\
L 199	&	 III$-$63	&	250.385576	&	36.411789	&	12.20	&	8.944	&	3.08	&	4095	&	0.65	&	$-$1.61	&	2.00	&	$-$1.66	&	$-$1.56	&	$+$0.10	&	$+$0.20	&	$-$247.7	\\
L 853	&	\nodata	&	250.447669	&	36.474579	&	12.20	&	8.927	&	3.09	&	4090	&	0.65	&	$-$1.51	&	2.00	&	$-$1.59	&	$-$1.43	&	$+$0.01	&	$+$0.20	&	$-$246.5	\\
L 261	&	\nodata	&	250.394873	&	36.466564	&	12.23	&	8.891	&	3.09	&	4050	&	0.65	&	$-$1.59	&	2.00	&	$-$1.66	&	$-$1.52	&	$+$0.00	&	$+$0.06	&	$-$233.9	\\
L 262	&	\nodata	&	250.395118	&	36.455494	&	12.25	&	9.000	&	3.06	&	4100	&	0.70	&	$-$1.58	&	2.00	&	$-$1.68	&	$-$1.48	&	$+$0.06	&	$+$0.22	&	$-$239.7	\\
L  72	&	 III$-$73	&	250.350357	&	36.425182	&	12.32	&	9.242	&	3.00	&	4000	&	0.65	&	$-$1.59	&	2.30	&	$-$1.79	&	$-$1.38	&	$+$0.14	&	$-$0.16	&	$-$243.0	\\
L 240	&	 II$-$34	&	250.391365	&	36.501595	&	12.34	&	9.088	&	3.03	&	4100	&	0.70	&	$-$1.42	&	2.00	&	$-$1.54	&	$-$1.30	&	$-$0.40	&	$+$0.30	&	$-$236.6	\\
L 481	&	\nodata	&	250.415993	&	36.474773	&	12.34	&	9.163	&	3.01	&	4145	&	0.75	&	$-$1.49	&	1.90	&	$-$1.56	&	$-$1.41	&	$-$0.77	&	$+$0.37	&	$-$238.1	\\
L 250	&	\nodata	&	250.393700	&	36.443489	&	12.37	&	9.211	&	2.99	&	4155	&	0.80	&	$-$1.58	&	1.95	&	$-$1.66	&	$-$1.49	&	$-$0.28	&	$+$0.38	&	$-$247.7	\\
L 384	&	\nodata	&	250.408606	&	36.463726	&	12.48	&	9.541	&	2.91	&	4295	&	0.90	&	$-$1.61	&	1.75	&	$-$1.53	&	$-$1.68	&	$+$0.36	&	$+$0.04	&	$-$243.1	\\
L 465	&	\nodata	&	250.415078	&	36.460598	&	12.50	&	9.583	&	2.89	&	4310	&	0.95	&	$-$1.64	&	1.90	&	$-$1.56	&	$-$1.71	&	$+$0.38	&	$-$0.15	&	$-$250.1	\\
L  96	&	 II$-$76	&	250.362885	&	36.466747	&	12.52	&	9.460	&	2.91	&	4215	&	0.90	&	$-$1.63	&	2.00	&	$-$1.76	&	$-$1.49	&	$+$0.43	&	$-$0.09	&	$-$236.3	\\
L 845	&	\nodata	&	250.446598	&	36.462189	&	12.53	&	9.499	&	2.90	&	4235	&	0.90	&	$-$1.57	&	1.85	&	$-$1.58	&	$-$1.56	&	$-$0.48	&	$+$0.40	&	$-$257.4	\\
L 745	&	 I$-$13	&	250.436927	&	36.514297	&	12.54	&	9.351	&	2.93	&	4135	&	0.85	&	$-$1.55	&	2.00	&	$-$1.66	&	$-$1.43	&	$+$0.20	&	$-$0.36	&	$-$242.7	\\
L 584	&	\nodata	&	250.423848	&	36.455242	&	12.55	&	9.857	&	2.83	&	4480	&	1.05	&	$-$1.63	&	1.85	&	$-$1.49	&	$-$1.77	&	$+$0.23	&	$+$0.32	&	$-$251.7	\\
L 296	&	\nodata	&	250.398768	&	36.446945	&	12.56	&	9.659	&	2.87	&	4320	&	0.95	&	$-$1.71	&	2.30	&	$-$1.78	&	$-$1.63	&	$+$0.26	&	$+$0.28	&	$-$246.6	\\
L 367	&	\nodata	&	250.406780	&	36.465542	&	12.56	&	9.575	&	2.88	&	4260	&	0.95	&	$-$1.57	&	1.85	&	$-$1.60	&	$-$1.53	&	$-$0.44	&	$+$0.40	&	$-$244.2	\\
L 316	&	 III$-$59	&	250.401230	&	36.419556	&	12.58	&	9.513	&	2.89	&	4210	&	0.90	&	$-$1.62	&	2.00	&	$-$1.65	&	$-$1.59	&	$-$0.18	&	$+$0.24	&	$-$238.7	\\
L 549	&	\nodata	&	250.421200	&	36.465279	&	12.58	&	9.996	&	2.80	&	4570	&	1.15	&	$-$1.62	&	1.75	&	$-$1.46	&	$-$1.78	&	$+$0.27	&	$+$0.44	&	$-$253.6	\\
L 674	&	\nodata	&	250.430845	&	36.454346	&	12.58	&	9.771	&	2.84	&	4390	&	1.05	&	$-$1.62	&	1.85	&	$-$1.58	&	$-$1.66	&	$+$0.01	&	$+$0.44	&	$-$244.2	\\
L 398	&	\nodata	&	250.409644	&	36.475033	&	12.60	&	9.775	&	2.84	&	4375	&	1.05	&	$-$1.55	&	1.85	&	$-$1.51	&	$-$1.59	&	$+$0.25	&	$+$0.08	&	$-$235.7	\\
L 244	&	 III$-$52	&	250.393224	&	36.418015	&	12.67	&	9.622	&	2.85	&	4220	&	0.95	&	$-$1.62	&	1.90	&	$-$1.66	&	$-$1.58	&	$-$0.08	&	$+$0.26	&	$-$245.8	\\
L 252	&	 II$-$33	&	250.393145	&	36.503727	&	12.67	&	9.643	&	2.85	&	4235	&	0.95	&	$-$1.60	&	1.90	&	$-$1.65	&	$-$1.55	&	$+$0.22	&	$+$0.02	&	$-$238.0	\\
L 830	&	\nodata	&	250.445026	&	36.451096	&	12.68	&	9.758	&	2.82	&	4305	&	1.00	&	$-$1.59	&	1.90	&	$-$1.59	&	$-$1.58	&	$-$0.32	&	$+$0.52	&	$-$247.6	\\
L 938	&	 IV$-$53	&	250.456950	&	36.452896	&	12.68	&	9.887	&	2.80	&	4400	&	1.10	&	$-$1.63	&	1.95	&	$-$1.58	&	$-$1.68	&	$+$0.46	&	$-$0.42	&	$-$239.7	\\
L 158	&	 II$-$57	&	250.377230	&	36.495411	&	12.70	&	9.727	&	2.82	&	4270	&	1.00	&	$-$1.58	&	1.90	&	$-$1.65	&	$-$1.51	&	$-$0.22	&	$+$0.06	&	$-$246.2	\\
L 666	&	\nodata	&	250.430458	&	36.440987	&	12.73	&	9.794	&	2.80	&	4295	&	1.00	&	$-$1.57	&	1.75	&	$-$1.54	&	$-$1.60	&	$+$0.27	&	$+$0.12	&	$-$251.7	\\
L  77	&	 III$-$18	&	250.352673	&	36.429165	&	12.77	&	9.830	&	2.79	&	4295	&	1.05	&	$-$1.62	&	1.95	&	$-$1.63	&	$-$1.60	&	$-$0.24	&	$+$0.44	&	$-$233.8	\\
L 169	&	 III$-$37	&	250.379465	&	36.432983	&	12.78	&	9.891	&	2.78	&	4330	&	1.05	&	$-$1.52	&	1.80	&	$-$1.55	&	$-$1.48	&	$-$0.34	&	$+$0.49	&	$-$246.7	\\
L 825	&	\nodata	&	250.444365	&	36.472321	&	12.78	&	10.011	&	2.75	&	4420	&	1.15	&	$-$1.55	&	1.80	&	$-$1.54	&	$-$1.55	&	$-$0.06	&	$+$0.43	&	$-$241.8	\\
L 353	&	 II$-$40	&	250.405233	&	36.493576	&	12.83	&	9.923	&	2.76	&	4315	&	1.10	&	$-$1.58	&	2.00	&	$-$1.67	&	$-$1.49	&	$+$0.35	&	$-$0.23	&	$-$237.2	\\
L 594	&	\nodata	&	250.424539	&	36.470486	&	12.86	&	10.099	&	2.72	&	4425	&	1.15	&	$-$1.57	&	1.80	&	$-$1.53	&	$-$1.60	&	$+$0.12	&	$+$0.45	&	$-$252.3	\\
L 777	&	 I$-$24	&	250.439463	&	36.499626	&	12.86	&	10.045	&	2.73	&	4385	&	1.15	&	$-$1.57	&	1.85	&	$-$1.58	&	$-$1.55	&	$+$0.20	&	$+$0.14	&	$-$244.6	\\
L 1073	&	\nodata	&	250.503562	&	36.392738	&	12.88	&	10.130	&	2.71	&	4430	&	1.15	&	$-$1.64	&	2.10	&	$-$1.72	&	$-$1.56	&	$+$0.32	&	$+$0.06	&	$-$250.9	\\
L 754	&	\nodata	&	250.438238	&	36.470348	&	12.88	&	10.058	&	2.72	&	4380	&	1.15	&	$-$1.60	&	1.80	&	$-$1.62	&	$-$1.58	&	$+$0.34	&	$+$0.09	&	$-$242.9	\\
L 198	&	\nodata	&	250.384773	&	36.459621	&	12.95	&	10.163	&	2.69	&	4405	&	1.20	&	$-$1.53	&	1.65	&	$-$1.47	&	$-$1.59	&	$+$0.02	&	$+$0.47	&	$-$247.6	\\
L 687	&	 IV$-$15	&	250.433291	&	36.376057	&	12.96	&	10.285	&	2.67	&	4490	&	1.25	&	$-$1.65	&	2.00	&	$-$1.73	&	$-$1.57	&	$+$0.10	&	$+$0.13	&	$-$251.0	\\
L 863	&	 I$-$42	&	250.448676	&	36.485596	&	12.98	&	10.299	&	2.66	&	4490	&	1.25	&	$-$1.59	&	1.80	&	$-$1.52	&	$-$1.66	&	$-$0.11	&	$+$0.11	&	$-$248.2	\\
L 1023	&	IV$-$61	&	250.475473	&	36.435909	&	12.99	&	10.285	&	2.66	&	4470	&	1.25	&	$-$1.50	&	1.75	&	$-$1.48	&	$-$1.51	&	$-$0.21	&	$+$0.31	&	$-$244.6	\\
L 877	&	 I$-$50	&	250.450037	&	36.498093	&	13.03	&	10.313	&	2.64	&	4460	&	1.25	&	$-$1.66	&	1.90	&	$-$1.64	&	$-$1.68	&	$+$0.21	&	$+$0.37	&	$-$245.7	\\
L 919	&	 IV$-$28	&	250.455285	&	36.413113	&	13.03	&	10.181	&	2.67	&	4360	&	1.20	&	$-$1.62	&	1.95	&	$-$1.62	&	\nodata	&	$-$0.13	&	$+$0.35	&	$-$255.6	\\
K 656	&	\nodata	&	250.561750	&	36.455540	&	13.04	&	10.191	&	2.66	&	4360	&	1.20	&	$-$1.57	&	1.85	&	$-$1.64	&	$-$1.49	&	$+$0.26	&	$+$0.03	&	$-$242.2	\\
L 343	&	\nodata	&	250.404518	&	36.453026	&	13.07	&	10.348	&	2.63	&	4455	&	1.25	&	$-$1.60	&	1.75	&	$-$1.56	&	$-$1.63	&	$+$0.35	&	$+$0.19	&	$-$241.8	\\
L 592	&	\nodata	&	250.424562	&	36.471794	&	13.10	&	10.423	&	2.61	&	4490	&	1.30	&	$-$1.68	&	1.85	&	$-$1.65	&	$-$1.70	&	$+$0.28	&	$+$0.31	&	$-$243.1	\\
L 476	&	\nodata	&	250.415991	&	36.448318	&	13.13	&	10.426	&	2.60	&	4470	&	1.30	&	$-$1.62	&	1.90	&	$-$1.60	&	$-$1.63	&	$+$0.30	&	$+$0.22	&	$-$230.1	\\
L 269	&	\nodata	&	250.395629	&	36.458511	&	13.14	&	10.457	&	2.60	&	4485	&	1.30	&	$-$1.65	&	1.75	&	$-$1.81	&	$-$1.48	&	$-$0.11	&	$+$0.21	&	$-$242.7	\\
L 948	&	 IV$-$35	&	250.458472	&	36.419598	&	13.14	&	10.559	&	2.58	&	4565	&	1.35	&	$-$1.64	&	1.90	&	$-$1.70	&	$-$1.58	&	$+$0.25	&	$+$0.18	&	$-$246.0	\\
L 967	&	 I$-$86	&	250.460928	&	36.471931	&	13.15	&	10.482	&	2.59	&	4500	&	1.35	&	$-$1.55	&	1.70	&	$-$1.52	&	$-$1.57	&	$-$0.46	&	$+$0.38	&	$-$256.0	\\
L 1030	&	I$-$77	&	250.477617	&	36.456387	&	13.18	&	10.482	&	2.58	&	4475	&	1.30	&	$-$1.61	&	1.75	&	$-$1.53	&	$-$1.68	&	$+$0.31	&	$+$0.15	&	$-$255.3	\\
L 644	&	\nodata	&	250.428298	&	36.483116	&	13.18	&	10.467	&	2.58	&	4460	&	1.30	&	$-$1.62	&	1.70	&	$-$1.57	&	$-$1.66	&	$-$0.04	&	$+$0.32	&	$-$233.1	\\
L 773	&	 I$-$23	&	250.439378	&	36.503170	&	13.21	&	10.554	&	2.56	&	4510	&	1.35	&	$-$1.66	&	1.90	&	$-$1.76	&	$-$1.56	&	$+$0.31	&	$+$0.34	&	$-$236.1	\\
L 956	&	\nodata	&	250.459008	&	36.471024	&	13.26	&	10.841	&	2.50	&	4720	&	1.50	&	$-$1.63	&	1.45	&	$-$1.67	&	$-$1.59	&	$+$0.18	&	$+$0.37	&	$-$242.6	\\
K 228	&	J 3	&	250.277032	&	36.470470	&	13.31	&	10.606	&	2.53	&	4470	&	1.35	&	$-$1.45	&	1.75	&	$-$1.52	&	$-$1.37	&	$-$0.06	&	$+$0.52	&	$-$249.0	\\
L 176	&	 II$-$87	&	250.380242	&	36.467278	&	13.31	&	10.702	&	2.51	&	4550	&	1.40	&	$-$1.61	&	1.75	&	$-$1.58	&	$-$1.63	&	$+$0.21	&	$+$0.37	&	$-$240.4	\\
K 188	&	 A1	&	250.179091	&	36.461632	&	13.39	&	10.704	&	2.50	&	4485	&	1.40	&	$-$1.52	&	1.70	&	$-$1.58	&	$-$1.46	&	$+$0.37	&	$+$0.22	&	$-$243.7	\\
L 436	&	\nodata	&	250.412601	&	36.443130	&	13.43	&	10.865	&	2.46	&	4580	&	1.50	&	$-$1.78	&	1.60	&	$-$1.76	&	$-$1.79	&	$+$0.08	&	$+$0.36	&	$-$233.0	\\
L 114	&	III$-$7	&	250.368221	&	36.451057	&	13.45	&	10.835	&	2.46	&	4540	&	1.45	&	$-$1.61	&	1.80	&	$-$1.61	&	$-$1.60	&	$+$0.16	&	$+$0.35	&	$-$250.6	\\
L 370	&	\nodata	&	250.407394	&	36.440029	&	13.47	&	10.917	&	2.44	&	4590	&	1.50	&	$-$1.66	&	1.60	&	$-$1.69	&	$-$1.63	&	$-$0.19	&	$+$0.27	&	$-$243.9	\\
L 1043	&	BAUM 13	&	250.480977	&	36.557415	&	13.49	&	11.229	&	2.39	&	4885	&	1.65	&	$-$1.67	&	1.80	&	$-$1.64	&	$-$1.70	&	$+$0.24	&	$+$0.57	&	$-$244.6	\\
L 766	&	 I$-$12	&	250.438708	&	36.518581	&	13.54	&	10.920	&	2.42	&	4540	&	1.50	&	$-$1.63	&	1.70	&	$-$1.63	&	$-$1.63	&	$-$0.02	&	$+$0.20	&	$-$247.9	\\
L 172	&	 III$-$45	&	250.380327	&	36.421734	&	13.56	&	11.008	&	2.41	&	4595	&	1.55	&	$-$1.58	&	1.60	&	$-$1.54	&	$-$1.61	&	$+$0.18	&	$+$0.17	&	$-$233.9	\\
L  26	&	 J38	&	250.320860	&	36.429989	&	13.58	&	10.991	&	2.40	&	4565	&	1.55	&	$-$1.51	&	1.60	&	$-$1.51	&	$-$1.50	&	$-$0.14	&	$+$0.28	&	$-$241.8	\\
L 168	&	 II$-$28	&	250.378456	&	36.503632	&	13.62	&	11.261	&	2.35	&	4780	&	1.65	&	$-$1.64	&	1.90	&	$-$1.67	&	$-$1.60	&	$+$0.61	&	$-$0.66	&	$-$245.6	\\
L 193	&	 II$-$94	&	250.382611	&	36.482044	&	13.63	&	11.139	&	2.37	&	4650	&	1.60	&	$-$1.56	&	1.70	&	$-$1.52	&	$-$1.60	&	$-$0.04	&	$+$0.55	&	$-$236.8	\\
L 726	&	 IV$-$19	&	250.436600	&	36.390942	&	13.68	&	11.104	&	2.36	&	4570	&	1.55	&	$-$1.54	&	1.55	&	$-$1.56	&	$-$1.52	&	$+$0.26	&	$+$0.16	&	$-$243.9	\\
L 793	&	\nodata	&	250.441763	&	36.451073	&	13.68	&	11.198	&	2.34	&	4660	&	1.65	&	$-$1.59	&	1.50	&	$-$1.53	&	$-$1.65	&	$+$0.54	&	$-$0.13	&	$-$248.5	\\
K 699	&	X 24	&	250.657823	&	36.403545	&	13.75	&	11.249	&	2.32	&	4640	&	1.65	&	$-$1.60	&	1.75	&	$-$1.62	&	$-$1.57	&	$+$0.22	&	$+$0.40	&	$-$249.3	\\
L 677	&	 IV$-$4	&	250.431585	&	36.429428	&	13.75	&	11.358	&	2.30	&	4750	&	1.70	&	$-$1.59	&	1.80	&	$-$1.48	&	$-$1.70	&	$-$0.06	&	$+$0.71	&	$-$250.2	\\
L  18	&	\nodata	&	250.313409	&	36.490002	&	13.78	&	11.206	&	2.32	&	4570	&	1.60	&	$-$1.65	&	1.80	&	$-$1.64	&	$-$1.66	&	$+$0.10	&	$+$0.11	&	$-$249.5	\\
L 800	&	 IV$-$18	&	250.443097	&	36.386150	&	13.80	&	11.295	&	2.30	&	4635	&	1.65	&	$-$1.59	&	1.70	&	$-$1.74	&	$-$1.44	&	$+$0.29	&	$+$0.24	&	$-$246.3	\\
L 1032	&	I$-$76	&	250.478956	&	36.461700	&	13.81	&	11.345	&	2.29	&	4600	&	1.60	&	$-$1.59	&	1.60	&	$-$1.69	&	$-$1.49	&	$-$0.19	&	$-$0.08	&	$-$246.0	\\
L 871	&	 I$-$19	&	250.449600	&	36.509624	&	13.89	&	11.396	&	2.26	&	4645	&	1.70	&	$-$1.62	&	1.70	&	$-$1.56	&	$-$1.67	&	$-$0.84	&	$+$0.63	&	$-$251.8	\\
L 955	&	 IV$-$22	&	250.459900	&	36.394932	&	13.92	&	11.382	&	2.26	&	4605	&	1.70	&	$-$1.54	&	1.55	&	$-$1.55	&	$-$1.52	&	$+$0.09	&	$+$0.29	&	$-$242.2	\\
L 609	&	\nodata	&	250.426025	&	36.439865	&	13.96	&	11.512	&	2.23	&	4690	&	1.75	&	$-$1.69	&	1.85	&	$-$1.67	&	$-$1.71	&	$+$0.39	&	$+$0.47	&	$-$251.2	\\
L  81	&	 II$-$23	&	250.353205	&	36.504230	&	13.97	&	11.474	&	2.23	&	4645	&	1.75	&	$-$1.60	&	1.60	&	$-$1.62	&	$-$1.57	&	$+$0.02	&	$+$0.56	&	$-$245.3	\\
L 1001	&	I$-$49	&	250.468308	&	36.477646	&	13.99	&	11.606	&	2.21	&	4755	&	1.80	&	$-$1.54	&	1.80	&	$-$1.53	&	$-$1.54	&	$+$0.24	&	$-$0.06	&	$-$257.8	\\
K 422	&	\nodata	&	250.401823	&	36.285603	&	14.02	&	11.568	&	2.21	&	4690	&	1.75	&	$-$1.54	&	1.65	&	$-$1.55	&	$-$1.52	&	$+$0.44	&	$+$0.22	&	$-$244.0	\\
L 1060	&	I$-$65	&	250.493449	&	36.475334	&	14.05	&	11.692	&	2.18	&	4785	&	1.85	&	$-$1.57	&	1.60	&	$-$1.56	&	$-$1.58	&	$-$0.08	&	$+$0.21	&	$-$236.5	\\
L 162	&	 III$-$43	&	250.378740	&	36.425434	&	14.09	&	11.603	&	2.18	&	4655	&	1.80	&	$-$1.48	&	1.40	&	$-$1.43	&	$-$1.52	&	$-$0.08	&	$+$0.08	&	$-$247.1	\\
L 488	&	\nodata	&	250.416136	&	36.488148	&	14.11	&	11.633	&	2.17	&	4665	&	1.80	&	$-$1.55	&	1.45	&	$-$1.63	&	$-$1.46	&	$+$0.20	&	$+$0.25	&	$-$243.0	\\
L 1051	&	IV$-$78	&	250.488983	&	36.422718	&	14.12	&	11.653	&	2.17	&	4675	&	1.80	&	$-$1.57	&	1.65	&	$-$1.64	&	$-$1.49	&	$-$0.29	&	$+$0.37	&	$-$243.8	\\
L 1114	&	\nodata	&	250.541079	&	36.373940	&	14.12	&	11.905	&	2.13	&	4935	&	1.95	&	$-$1.65	&	1.65	&	$-$1.71	&	$-$1.59	&	$+$0.30	&	$+$0.01	&	$-$255.4	\\
L 557	&	\nodata	&	250.420067	&	36.569859	&	14.13	&	11.657	&	2.16	&	4670	&	1.80	&	$-$1.58	&	1.50	&	$-$1.61	&	$-$1.55	&	$+$0.45	&	$-$0.21	&	$-$247.9	\\
L 79	&	\nodata	&	250.352099	&	36.493851	&	14.18	&	11.809	&	2.13	&	4770	&	1.90	&	$-$1.55	&	1.40	&	$-$1.51	&	$-$1.59	&	$+$0.55	&	$-$0.26	&	$-$242.4	\\
K 659	&	\nodata	&	250.568872	&	36.416309	&	14.24	&	11.801	&	2.12	&	4700	&	1.85	&	$-$1.39	&	1.50	&	$-$1.47	&	$-$1.30	&	$-$0.01	&	$+$0.33	&	$-$240.9	\\
L 140	&	 III$-$25	&	250.375557	&	36.443192	&	14.24	&	12.242	&	2.05	&	5195	&	2.50	&	$-$1.61	&	1.50	&	$-$1.57	&	$-$1.64	&	$+$0.11	&	$+$0.61	&	$-$237.2	\\
K 674	&	\nodata	&	250.588623	&	36.564819	&	14.26	&	12.147	&	2.06	&	5055	&	2.05	&	$-$1.59	&	1.80	&	$-$1.61	&	$-$1.56	&	$+$0.69	&	$+$0.26	&	$-$249.0	\\
L 787	&	 I$-$2	&	250.440208	&	36.529823	&	14.26	&	11.903	&	2.09	&	4785	&	1.90	&	$-$1.58	&	1.65	&	$-$1.53	&	$-$1.63	&	$+$0.30	&	$+$0.47	&	$-$246.2	\\
L 1050	&	\nodata	&	250.489252	&	36.387623	&	14.29	&	11.893	&	2.09	&	4745	&	1.90	&	$-$1.47	&	1.70	&	$-$1.47	&	$-$1.46	&	$+$0.07	&	\nodata	&	$-$245.7	\\
L 1072	&	IV$-$80	&	250.501614	&	36.423615	&	14.31	&	11.973	&	2.07	&	4805	&	1.95	&	$-$1.55	&	1.70	&	$-$1.57	&	$-$1.52	&	$+$0.10	&	$+$0.64	&	$-$244.9	\\
L 423	&	 II$-$7	&	250.410956	&	36.510258	&	14.31	&	11.938	&	2.08	&	4770	&	1.95	&	$-$1.51	&	1.50	&	$-$1.48	&	$-$1.54	&	$+$0.51	&	$-$0.11	&	$-$237.1	\\
L 824	&	 I$-$39	&	250.444099	&	36.490067	&	14.34	&	12.094	&	2.05	&	4905	&	2.00	&	$-$1.56	&	1.70	&	$-$1.46	&	$-$1.65	&	$+$0.31	&	$+$0.37	&	$-$240.4	\\
L 1096	&	I$-$67	&	250.513947	&	36.486725	&	14.40	&	12.048	&	2.04	&	4790	&	2.00	&	$-$1.58	&	1.40	&	$-$1.58	&	\nodata	&	$-$0.07	&	$+$0.33	&	$-$250.9	\\
L 1097	&	\nodata	&	250.514202	&	36.505070	&	14.40	&	12.061	&	2.04	&	4805	&	2.00	&	$-$1.48	&	1.55	&	$-$1.50	&	$-$1.46	&	$-$0.32	&	$+$0.30	&	$-$256.2	\\
L  29	&	 II$-$63	&	250.323850	&	36.491638	&	14.51	&	12.186	&	1.99	&	4820	&	2.05	&	$-$1.59	&	1.80	&	$-$1.60	&	$-$1.58	&	\nodata	&	$+$0.52	&	$-$237.9	\\
K 224	&	J 37	&	250.275644	&	36.422974	&	14.52	&	12.225	&	1.98	&	4850	&	2.05	&	$-$1.53	&	1.55	&	$-$1.51	&	$-$1.54	&	$+$0.38	&	$+$0.50	&	$-$246.3	\\
L 137	&	\nodata	&	250.373757	&	36.537121	&	14.58	&	12.315	&	1.95	&	4880	&	2.10	&	$-$1.47	&	1.30	&	$-$1.42	&	$-$1.52	&	$+$0.22	&	$+$0.40	&	$-$245.1	\\
L  16	&	 J50	&	250.312979	&	36.398304	&	14.60	&	12.388	&	1.94	&	4940	&	2.15	&	$-$1.61	&	1.70	&	$-$1.53	&	$-$1.69	&	$+$0.44	&	$-$0.18	&	$-$245.3	\\
K 647	&	\nodata	&	250.546518	&	36.306267	&	14.71	&	12.488	&	1.89	&	4930	&	2.20	&	$-$1.58	&	1.70	&	$-$1.61	&	$-$1.55	&	$-$0.07	&	$+$0.07	&	$-$244.0	\\
L 93	&	III$-$40	&	250.361839	&	36.424629	&	14.85	&	12.654	&	1.83	&	4960	&	2.25	&	$-$1.42	&	2.10	&	$-$1.35	&	$-$1.49	&	$+$0.42	&	$+$0.39	&	$-$245.8	\\
L 1095	&	I$-$69	&	250.513806	&	36.465366	&	14.92	&	12.663	&	1.82	&	4400	&	1.30	&	$-$1.62	&	1.90	&	$-$1.62	&	$-$1.61	&	$+$0.32	&	$-$0.06	&	$-$242.0	\\
L   6	&	 J11	&	250.291419	&	36.502876	&	15.00	&	14.352	&	1.70	&	4450	&	1.30	&	$-$1.68	&	1.85	&	$-$1.74	&	$-$1.62	&	$-$0.02	&	$+$0.36	&	$-$251.3	\\
L 101	&	II$-$60	&	250.363988	&	36.491802	&	15.09	&	13.027	&	1.72	&	4500	&	1.40	&	$-$1.68	&	1.80	&	$-$1.66	&	$-$1.69	&	$+$0.40	&	$-$0.32	&	$-$243.6	\\
L  32	&	 II$-$64	&	250.327637	&	36.478725	&	15.12	&	12.937	&	1.73	&	4850	&	2.10	&	$-$1.44	&	2.00	&	$-$1.34	&	$-$1.54	&	$+$0.01	&	$+$0.50	&	$-$242.3	\\
CM 12	&	\nodata	&	250.556994	&	36.554123	&	15.28	&	13.122	&	1.66	&	5000	&	2.45	&	$-$1.39	&	1.70	&	$-$1.28	&	$-$1.50	&	$+$0.74	&	$-$0.01	&	$-$235.5	\\
\enddata

\tablenotetext{a}{Designations from Ludendorf (1905) and Kadla (1966).}

\end{deluxetable}